\documentclass[review]{elsarticle}

\usepackage{hyperref}

\makeatletter
\def\ps@pprintTitle{%
 \let\@oddhead\@empty
 \let\@evenhead\@empty
 \def\@oddfoot{\centerline{\thepage}}%
 \let\@evenfoot\@oddfoot}
\makeatother

\usepackage{numcompress}

\biboptions{sort&compress}

\begin{document}

\begin{frontmatter}

\title{A simple parametrisation for coupled dark energy}


\author[IA]{Vitor da Fonseca}
\author[IA,ECEO]{Tiago Barreiro}
\author[IA]{Nelson J. Nunes}

\address[IA]{Instituto de Astrofísica e Ciências do Espaço, \\Faculdade de Ciências da Universidade de Lisboa,\\Campo Grande, PT1749-016 Lisboa, Portugal}
\address[ECEO]{ECEO, Universidade Lusófona de Humanidades e Tecnologias, \\Campo Grande, 376, 1749-024 Lisboa, Portugal}

\begin{abstract}
As an alternative to the popular parametrisations of the dark energy equation of state, we construct a quintessence model where the scalar field  has a linear dependence on the number of e-folds.  Constraints on more complex models are typically limited by  the degeneracies that increase with the number of parameters. The proposed parametrisation conveniently constrains the evolution of the dark energy equation of state as it allows for a wide variety of time evolutions. We also consider a non-minimal coupling to cold dark matter. We fit the model with Planck and KiDS observational data. The CMB favours a non-vanishing coupling with energy transfer from dark energy to dark matter. Conversely, gravitational weak lensing measurements slightly favour energy transfer from dark matter to dark energy, with a substantial departure of the dark energy equation of state from -1.
\end{abstract}


\end{frontmatter}

\section{\label{sec:intro}Introduction}

The accelerated expansion of the Universe was discovered by two independent teams who surprisingly found in 1998 that remote type Ia supernovae were fainter than expected in a universe presumably dominated by matter \cite{acel1,acel2}. This pivotal discovery suggests the existence of a repulsive component, termed dark energy, filling the cosmos with a negative pressure that counterbalances gravitational attraction, provided that both general relativity and the cosmological principle still hold \cite{amendola,mod}. The cosmological constant $\Lambda$ that represents the energy of vacuum is a straightforward and natural candidate for dark energy \cite{cc1}. While the $\Lambda$CDM model is the simplest one in agreement with astrophysical observations \cite{collaboration2018planck}, it suffers from several shortcomings \cite{martin}, notably the coincidence problem that can be addressed by considering that dark energy varies in time \cite{scalar3}. In this respect, many dark energy models beyond the cosmological constant have been investigated, as well as alternatives in the form of modified gravity on large scales \cite{2006copeland,Brax_2017,saridakis2021modified}. The simplest and model independent approach is to phenomenologically parametrise in redshift the dark energy equation of state by ideally introducing the fewest possible parameters to limit degeneracies \cite{param_z,param_a1,param_a2,param_log,Corasaniti}. The detection of any time dependence would be a crucial information excluding a cosmological constant. This approach, albeit very common, is usually restricted to the local Universe.

Instead, we propose with a reduced number of parameters to parametrise the evolution of a quintessence scalar field $\phi$ supposedly responsible for a dynamical dark energy component both at the early and present times \cite{Wetterich:1994bg}. Given the enigmatic nature of dark energy, it is logical to look into known physics to at least attempt to describe its possible dynamics. Scalar fields are promising candidates because they already exist in particle physics models \cite{higgs}. The scalar field parametrisation we contemplate is sufficiently simple to be constrained by observations. It takes the form of a linear dependence on the number of e-folds, such that $\phi^\prime=\lambda$ \cite{Nunes:2003ff} for some constant $\lambda$, where the prime denotes the derivative with respect to the number of e-folds. Furthermore, the unknown nature of both dark energy and dark matter allows models that assume an interaction between them \cite{PhysRevD.62.043511}. Interacting dark energy potentially addresses two of the current open cosmological questions posed by the concordance model \cite{2020valentino,2019martinelli,2021nunes,Sol_Peracaula_2021}. It tends to alleviate the tension between the rates of expansion $H_0$ ascertained by early and late Universe probes \cite{2019verde}, as well as the $\sigma_8$ tension between large-scale structure observations and those from the Planck satellite.  For the sake of simplicity, we assume here that the interaction between quintessence and dark matter is of constant strength, $\beta$ \cite{article_barros}, corresponding to a conformal coupling that appears in the Einstein frame where the matter fields depend on the scalar field \cite{Tsujikawa_2008}. 

We begin in Section~\ref{sec:parametrisation} with the description of the parametrisation envisioned and we examine its behaviour in terms of background cosmology. In Section~\ref{sec:pertubations}, we explore the impact of the parametrisation on the cosmological evolution of the perturbations and we present the analytic expressions we derive for the dark matter fluctuations in the Newtonian limit. In Section~\ref{sec:constraints}, we extract constraints on the parameters, based on observations of the cosmic microwave background (CMB) anisotropies by the Planck satellite, as well as weak gravitational lensing measurements from KiDS-450 probing large scale structures. Finally, Section~\ref{sec:discussion} discusses the conflicting results we discover on the constraints between the two sets of observations. 
\section{\label{sec:parametrisation}Background cosmology with a parametrised scalar field}
We assume  dark energy in the form of quintessence, i.e. a canonical and homogeneous scalar field $\phi$ responsible for the late time acceleration of the Universe expansion \cite{Caldwell_1998,tsuq}. In such theory, the Lagrangian density of the scalar field sums a potential energy term to the standard kinetic energy one,
\begin{equation}
{\cal{L}_\phi} = -\frac{1}{2}g^{\mu\nu}\partial_\mu\phi\partial_\nu\phi-V(\phi),
\label{eq:lagrangian}
\end{equation}
where $g_{\mu\nu}$ is the metric and $V\left(\phi\right)$ is an undetermined function of $\phi$ that defines the self-interacting potential of quintessence. We suppose that the geometry of the cosmological background is homogenous and isotropic, described by a flat Friedmann-Lemaître-Robertson-Walker (FLRW) spacetime whose line element reads in cosmic time,
\begin{equation}
ds^2=-dt^2+a^2(t)\delta_{ij}dx^idx^j,
\label{eq:line_element}
\end{equation}
where $a$ is the Universe scale factor. Besides dark energy, we assume that the Universe is composed of radiation $(r)$, baryons $(b)$ and cold dark matter $(c)$. In that case, the energy densities $\rho_i$ of the different species $i$ satisfy the Friedmann constraint,
\begin{equation}
H^2=\frac{\kappa^2}{3}\left(\rho_r+\rho_b+\rho_c+\rho_\phi\right),
\label{eq:friedmann}
\end{equation}
where $\kappa^2\equiv8\pi G$ ($G$ is the gravitational constant), $H\equiv\dot{a}/a$ is the Hubble expansion rate, $\rho_\phi=\dot{\phi}^2/2+V\left(\phi\right)$ represents the energy density of the scalar field approximated by a perfect fluid, the dot denoting differentiation with respect to cosmic time.

We also consider that besides being minimally coupled to gravity, the scalar field is non-minimally coupled to the dark matter component through a constant interaction \cite{PhysRevD.62.043511}. Within the dark sector, the energy-momentum tensor is jointly conserved according to the Bianchi identities, preserving the covariance of the theory,
\begin{equation}
\label{eq:conservation_total}
\nabla_\mu\left(T^{(c)}\,^{\mu}_{\nu}+T^{(\phi)}\,^{\mu}_{\nu}\right)=0,
\end{equation}
however, the energy-momentum tensors are not individually conserved,
\begin{eqnarray}
\label{eq:conservation:1}
\nabla_\mu T^{(c)}\,^{\mu}_{\nu}&=+\kappa\beta T_c\nabla_\mu\phi, \\
\label{eq:conservation:2}
\nabla_\mu T^{(\phi)}\,^{\mu}_{\nu}&=-\kappa\beta T_c\nabla_\mu\phi,
\end{eqnarray}
where $\beta$ is the constant strength of the coupling, and $T_c=-\rho_c+3p_c=-\rho_c$ is the trace of the energy-momentum tensor of dark matter, which is pressureless by definition. We conventionally set $\kappa=1$ for the rest of the document. The time component ($\nu=0$) of Eq.~(\ref{eq:conservation:1}) yields the evolution equation of dark matter,
\begin{equation}
\rho_c^\prime+3\rho_c=\beta\rho_c\phi^\prime\quad\Rightarrow\quad\rho_c=\rho_{c_0}~e^{-3N+\beta\left(\phi-\phi_0\right)}, \label{eq:cdm}
\end{equation}
and the time component of Eq.~(\ref{eq:conservation:2}) gives  the evolution equation of quintessence,
\begin{equation}
\rho_\phi^\prime+3H^2\phi^{\prime 2}=-\beta\rho_c\phi^\prime, \label{eq:phi}
\end{equation}
where a prime stands for the differentiation with respect to the number of e-folds, $N\equiv\ln a$, considered as the time variable. The latter equation can be rewritten as the so-called Klein-Gordon equation of motion,
\begin{equation}
\ddot{\phi}+3H\dot{\phi}+V_{,\phi}=-\beta\rho_c,
\label{eq:Klein_Gordon}
\end{equation}
where the subscript $,\phi$ stands for the derivative with respect to the scalar field. As for the other two cosmic fluid components, a similar coupling with radiation would vanish since the trace of its energy-momentum tensor is null, and we disregard the interactions with baryonic matter because of local gravity constraints. They are therefore conserved separately and evolve according to regular continuity equations,
\begin{eqnarray}
\label{eq:radiation}
\rho_r^\prime+4\rho_r=0\quad &\Rightarrow\quad\rho_r=\rho_{r_0}e^{-4N}, \\
\label{eq:baryons}
\rho_b^\prime+3\rho_b=0\quad &\Rightarrow\quad\rho_b=\rho_{b_0}e^{-3N}.
\end{eqnarray}

By substituting Eqs.~(\ref{eq:baryons}) and (\ref{eq:cdm}) into the Friedmann constraint (\ref{eq:friedmann}), we find that Eq.~(\ref{eq:phi}) can be rewritten as,
\begin{equation}
\rho_\phi^\prime+\phi^{\prime^2}\rho_\phi=-\left(\phi^{\prime^2}+\beta\phi^\prime\right)\rho_{c_0}~e^{-3N+\beta\left(\phi-\phi_0\right)}-\phi^{\prime^2}\rho_{b_0}~e^{-3N},
\label{eq:diff_phi}
\end{equation}
where we have neglected the radiation component. The general solution for the quintessence energy density is hence, 
\begin{equation}
\rho_\phi=e^{-\int_0^N dx\phi^{\prime^2}}\left\lbrace\rho_{\phi_0}-\int_0^Ndx\left[\left(\phi^{\prime^2}+\beta\phi^\prime\right)\rho_{c_0}e^{-3x+\beta\left(\phi-\phi_0\right)}+\phi^{\prime^2}\rho_{b_0}e^{-3x}\right]e^{\int_0^x dy\phi^{\prime^2}}\right\rbrace.
\label{eq:general_diff_phi}
\end{equation}

Following Ref.~\cite{Nunes:2003ff}, in order to obtain an analytic expression for the solution to Eq.~(\ref{eq:general_diff_phi}), we parametrise the evolution of the scalar field in the form of a linear function of the number of e-folds, with a given constant $\lambda$,
 \begin{equation}
\phi-\phi_0=\lambda N
\quad\Rightarrow\quad
\phi^\prime=\lambda.
\label{eq:param_phi}
\end{equation}
With such parametrisation, Eq.~(\ref{eq:general_diff_phi}) simplifies to,
\begin{eqnarray}
\frac{\rho_\phi}{3H_0^2}=&\left(1+\frac{3}{\lambda^2+\beta\lambda-3}~\Omega_{c_0}+\frac{3}{\lambda^2-3}~\Omega_{b_0}\right)e^{-\lambda^2N}\,
\label{eq:solution_diff_phi}
 \\
&-\frac{\lambda^2+\beta\lambda}{\lambda^2+\beta\lambda-3}~\Omega_{c_0}~e^{(-3+\beta\lambda)N}
-\frac{\lambda^2}{\lambda^2-3}~\Omega_{b_0}~e^{-3N}. \nonumber
\end{eqnarray}
where $H_0$ is today's expansion rate and $\Omega_{i_0}\equiv\rho_{i_0}/3H_0^2$ today's abundance parameter of species $i$.

Usually in the literature, the analytic form of the scalar field potential is freely and explicitly assumed a priori, along with the requirement that the field is slow-rolling in order to produce cosmic acceleration \cite{PhysRevLett.84.2076, PhysRevLett.85.5276, PhysRevD.62.103517, Martin:2008qp, scalar1, scalar2, PhysRevD.85.023503}. The latter assumption implies a flat potential in order to provide for a negative pressure of the cosmological fluid when the scalar field dominates the matter content of the Universe. However, here, we do not assume any prior slow-rolling conditions since the potential stems from the choice of our parametrisation. It can be found by noticing that,
\begin{equation}
V(\phi) = \rho_\phi-\frac{1}{2}H^2\lambda^2,
\end{equation}
which brings us to,
\begin{equation}
V(\phi)=Ae^{\left(-\frac{3}{\lambda}+\beta\right)\phi}+Be^{-\lambda\phi}+Ce^{-\frac{3}{\lambda}\phi},
\label{eq:potential}
\end{equation}
where the mass scales are given by,
\begin{eqnarray}
&A=&\frac{3}{2}\,\frac{\lambda^2+2\beta\lambda}{3-\lambda^2-\beta
\lambda}H_0^2\Omega_{c_0}e^{\left(\frac{3}{\lambda}-\beta\right)\phi_0},\nonumber\\
&B=&\frac{6-\lambda^2}{2}H_0^2\left(1+\frac{3}{\lambda^2-3+\beta\lambda}\Omega_{c_0}+\frac{3}{\lambda^2-3}\Omega_{b_0}\right)e^{\lambda\phi_0},
\nonumber\\
&C=&\frac{3}{2}\,\frac{\lambda^2}{3-\lambda^2}~H_0^2\Omega_{b_0}e^{\frac{3}{\lambda}\phi_0}.\nonumber
\end{eqnarray}
In the absence of interaction, $\beta=0$, we recover the potential reconstructed in Ref.~\cite{Nunes:2003ff}.

Denoting $p_\phi=\dot{\phi}^2/2-V\left(\phi\right)$ the pressure of the scalar field, the quintessence equation of state, $w_\phi=p_\phi/\rho_\phi$, satisfies,
\begin{equation}
\rho_\phi(1+w_\phi)=H^2\lambda^2,
\label{eq:p_phi}
\end{equation}
and therefore, in the absence of radiation,
\begin{equation}
w_\phi=-1+\frac{\lambda^2}{3}\left[1+\frac{3H_0^2}{\rho_\phi}\left(\Omega_{c_0}~e^{(-3+\beta\lambda)N}+\Omega_{b_0}~e^{-3N}\right)\right],
\label{eq:w_phi_coupled}
\end{equation}
where $3H_0^2/\rho_\phi$ is given by Eq.~(\ref{eq:solution_diff_phi}). Note that today's equation of state $w_{_0}$ is not affected by the coupling,
\begin{equation}
w_{_0}=-1+\frac{\lambda^2}{3\Omega_{\phi_0}}.
\end{equation}
Since we have evidence that the Universe is already accelerating, the current effective equation of state is bound by $w_{\textrm{eff}}=w_{_0}\Omega_{\phi_0}<-1/3$, still neglecting radiation. This ceiling imposes a condition applying to the parameter $\lambda$ that singly governs at the present time the evolution of the dynamical system composed of the Klein-Gordon and Friedmann equations \cite{Barreiro:1999zs,Amendola_2002},
\begin{equation}
|\lambda|<\sqrt{-1+3\Omega_{\phi_0}}<\sqrt{2},
\label{eq:lambda_acceleration}
\end{equation}
and consequently $\Omega_{\phi_0}>1/3$.
\begin{figure}[t]
\centering
\includegraphics[scale=0.4]{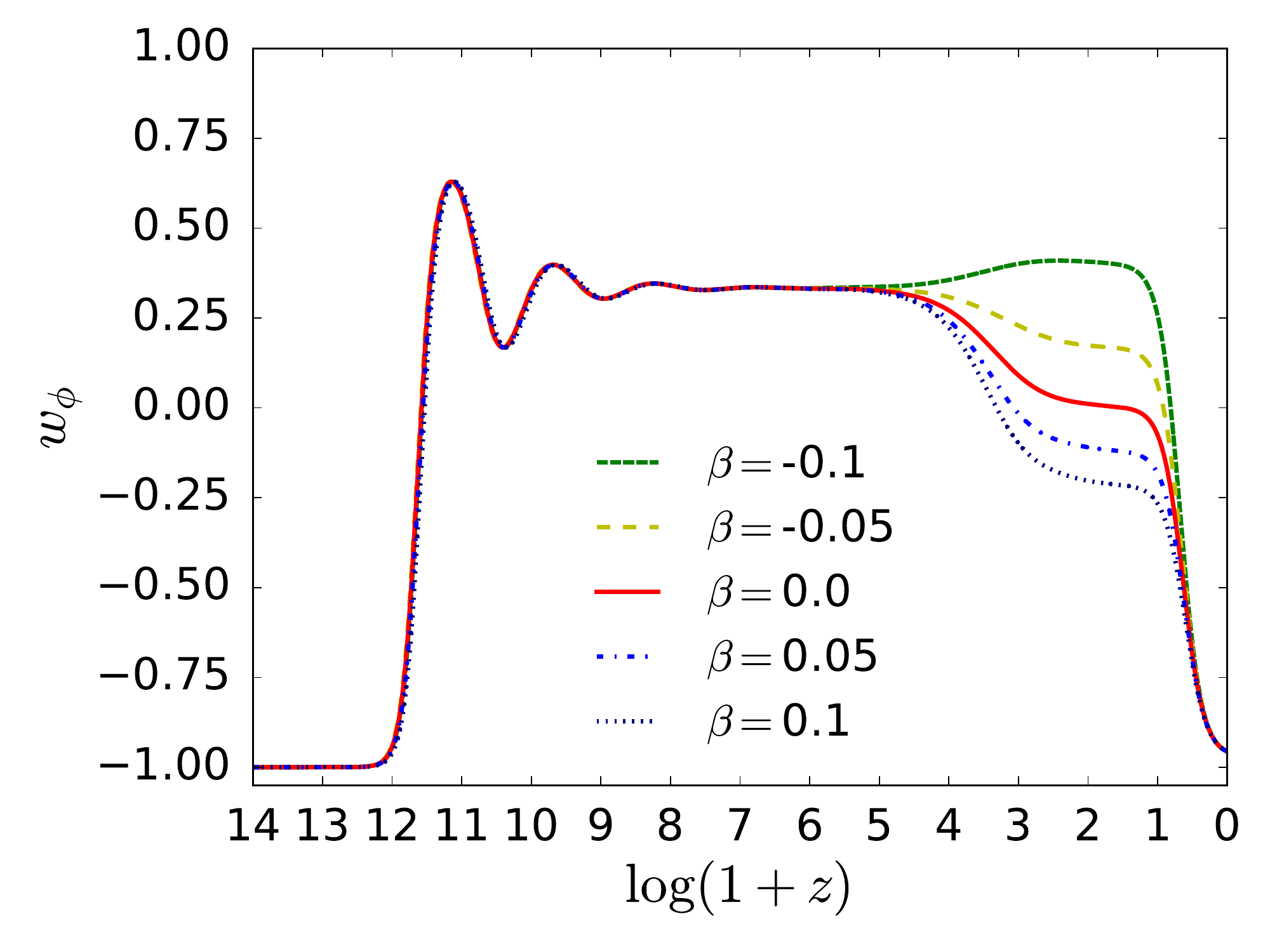}
  \caption{Dark energy equation of state for $\lambda=0.3$. It is frozen at the outset of the evolution and then oscillates during a transitory period prior reaching the radiation attractor. The coupling turns on during matter domination and turns off when the scalar field starts slow-rolling.}
\label{fig:w_phi}
\end{figure}

\begin{figure}[h]
\centering
\includegraphics[scale=0.4]{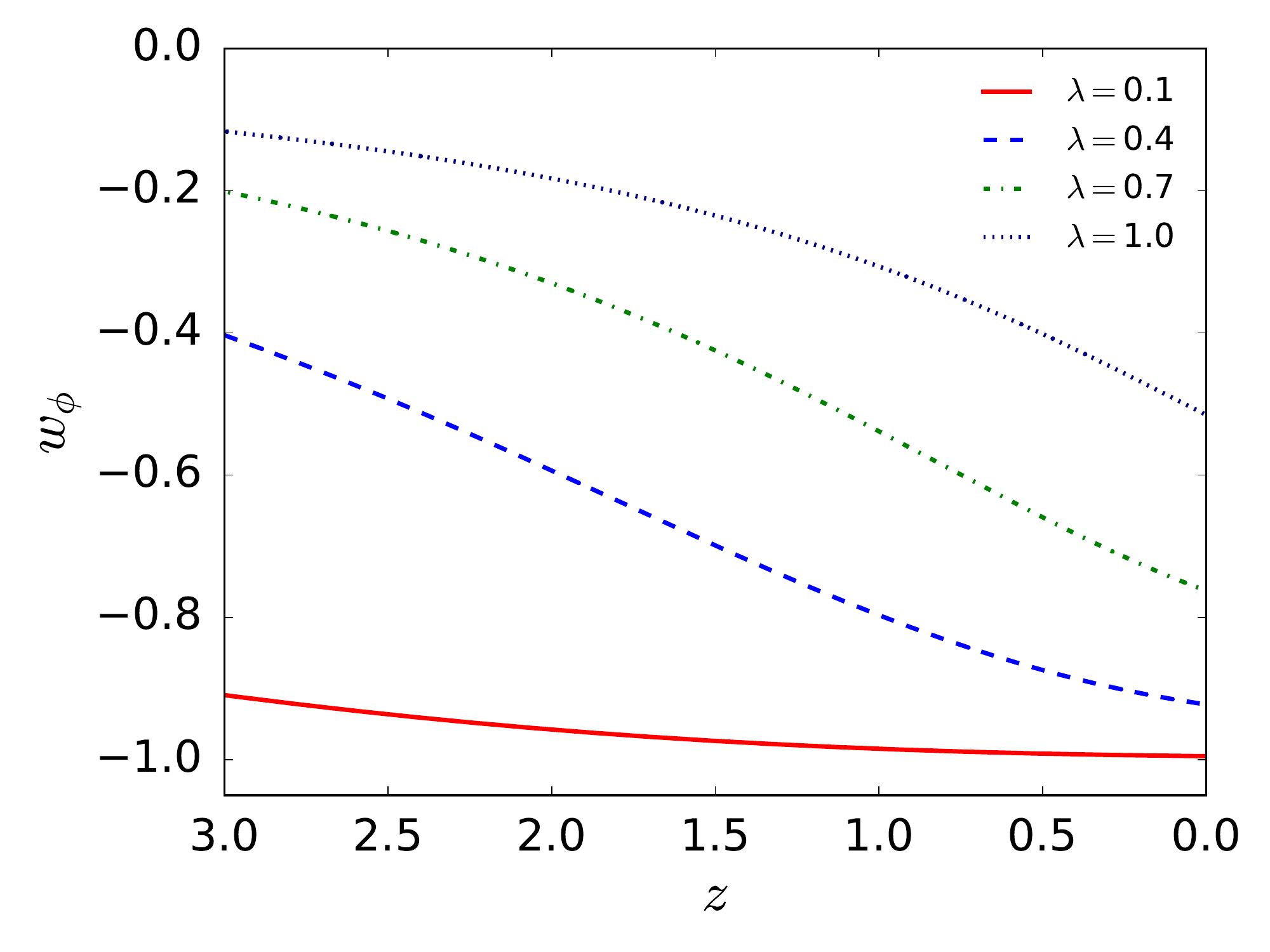}
\caption{\label{fig:low_redshift}Dark energy equation of state evolution at low redshift in the absence of coupling, $\beta=0$.}
\end{figure}

If one further neglects the baryons in the expression of the potential (\ref{eq:potential}), it is possible to analytically describe the cosmological evolution of $w_\phi$ following Ref.~\cite{amendola}. At early times, the potential is driven by the steeper exponential term $e^{(-3/\lambda+\beta)\phi}$ and quintessence provisionally scales with radiation,
\begin{equation}
w_\phi=\frac{1}{3},
\qquad
\Omega_\phi=\frac{4}{\left(\frac{3}{\lambda}-\beta\right)^2}.
\label{eq:scaling_radiation}
\end{equation}
It is subsequently attracted to a stable scaling solution in the matter era during which $w_\phi$ and $\Omega_\phi$ are also constant,
\begin{equation}
w_\phi=-\frac{\beta}{\lambda+\beta},
\qquad
\Omega_\phi=\frac{\lambda}{3}\left(\beta+\lambda\right).
\label{eq:scaling_matter}
\end{equation}
Later, once the shallower exponential term $e^{-\lambda\phi}$ takes the lead in driving the scalar field, dark energy density freezes and accelerates the Universe, independently of the coupling. The viability of this cosmological scenario is confirmed numerically in Fig.~\ref{fig:w_phi}. The dark energy equation of state $w_\phi$ is well behaved at high redshift as it is bound within the $[-1,1]$ interval by following a common attractor irrespectively of the parameters values.  Interestingly, at low redshift, the model catches a large span of possible evolutions that is determined by one single parameter, $\lambda$, as illustrated in Fig.~\ref{fig:low_redshift}.

\begin{figure}[tb]
\centering
\includegraphics[scale=0.4]{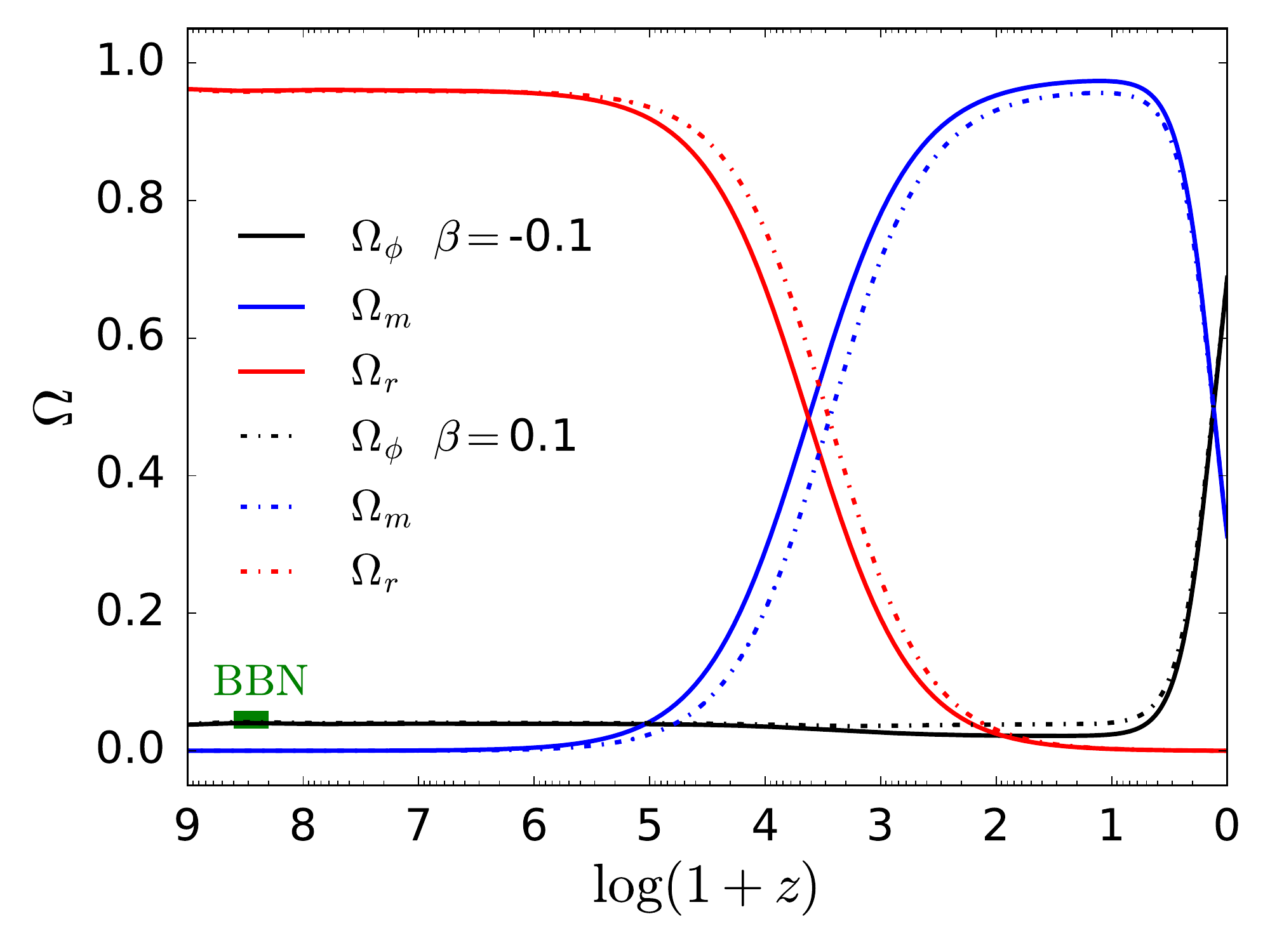}
  \caption{Background evolution of the fractional energy densities ($\lambda=0.3$). Early dark energy is capped by BBN. The horizontal green segment corresponds to the tighter bound in Eq.~(\ref{eq:BBN_conservative}).}
\label{fig:omega_coupled}
\end{figure}

There exists another theoretical bound on $\lambda$ due to a ceiling applying to the early dark energy abundance as depicted in Fig.~\ref{fig:omega_coupled}. The fractional energy density of quintessence at the Big Bang Nucleosynthesis (BBN), around $z\sim4\times10^8$, when the temperature of the Universe is about $1$ MeV, is capped by $\Omega_\phi<0.045$ at $2\sigma$ \cite{PhysRevD.64.103508}. Otherwise, the resulting production of primordial elements contradicts the observation of their current abundance. When the radiation attractor is reached before primordial nucleosynthesis\footnote{This is the attractor for the potential in Eq.~(\ref{eq:potential}). One could also opt to construct a potential imposing the scaling of $\phi^\prime=\lambda$ to hold during radiation too, by adding one extra exponential term with a $4/\lambda$ slope. This would give us the looser bound $|\lambda| < 0.43$.}, we find the following ceiling by employing Eq.~(\ref{eq:scaling_radiation}) where we neglect the value of the coupling $\beta$ before $3/\lambda$,
\begin{equation}
\Omega_\phi=\frac{4}{9}\lambda^2<0.045\quad\Rightarrow\quad|\lambda|<0.32.
\label{eq:BBN_conservative}
\end{equation}
The more conservative bounds $\Omega_\phi<0.13$--$0.2$ in Ref.~\cite{scalar3} respectively relax the constraints to $|\lambda|<0.54$ and $|\lambda|<0.67$.

The numerical results we present are obtained with a version of the Einstein-Boltzmann code CLASS \cite{Class1,Class2} that we modify to accommodate the coupled dark matter conservation in Eq.~(\ref{eq:cdm}) and the potential in Eq.~(\ref{eq:potential}), as well as the Klein-Gordon equation (\ref{eq:Klein_Gordon}). We vary our two parameters in the simulations while the other cosmological parameters are set at the values of the Planck 2018 results \cite{collaboration2018planck}. Moreover, since it is always possible to redefine the potential mass scales with today's value $\phi_{_0}$, we decide to set it to zero. We also slightly tune the initial conditions of the scalar field $\phi_i$ at redshift $z\equiv1/a-1=10^{14}$ in order to ensure that dark energy is sub-dominant at the outset of the evolution. Our approach is consistent with Ref.~\cite{PhysRevD.58.023503} where the initial conditions are set in the aftermath of inflation. The precise initial conditions are unimportant thanks to the tracking solutions. For the same reason, we decide to set $\dot{\phi}_i=0$.

Finally, we explore the behaviour of the parametrisation for $\lambda>0$ only, given that the equations are invariant under the transformations $\lambda\rightarrow-\lambda$, $\beta\rightarrow-\beta$ and $\phi\rightarrow-\phi$. The direction of the stress-energy flow inside the dark sector depends upon both the evolution of the scalar field imposed by the sign of $\phi^\prime=\lambda$ and the sign of the coupling $\beta$. In this specific case, the coupling pumps energy from the scalar field component into the dark matter fluid when $\beta>0$. Consequently, the energy being granted to dark matter slows its dilution. When $\beta<0$, the energy lost by dark matter to the benefit of the scalar field conversely accelerates dark matter's dilution.
\section{\label{sec:pertubations}Impact on the evolution of the perturbations}
The evolution of the primordial fluctuations of the cosmological fluid in an expanding FLRW universe can be studied with the linear perturbation theory as long as the perturbations to the homogeneous background are small enough \cite{Lifshitz:1945du}.

Like any physical quantity, the spacetime metric is expanded into spatial average and small linear perturbations. In the synchronous gauge, the scalar component of the inhomogeneous line element reduces to,
\begin{equation}
ds^2= a^2(\tau)\left[-d\tau^2+\left(\delta_{ij}+h_{ij}\right)dx^idx^j\right],
\end{equation}
in conformal time $d\tau\equiv dt/a$, where $h_{ij}$ is the rank-2 symmetric tensor field that corresponds to the six components of the perturbations spatial part. $h_{ij}$ is parametrised in Fourier space with two fields $h(\vec{k},\tau)$ and $\eta(\vec{k},\tau)$ as in Ref.~\cite{Ma_1995}. For the perturbations in the cosmological fluid, the dimensionless density contrast $\delta_i$ conveniently describes the fluctuations in the energy density field of a given cosmological species $i$,
\begin{equation}
\delta_i\equiv \frac{\delta\rho_i}{\bar{\rho_i}}=\frac{\rho_i(\vec{x},t)-\bar{\rho}_i(t)}{\bar{\rho}_i(t)},
\end{equation}
where the bar denotes the background quantities. Similarly, noting $\varphi=\delta\phi$, the perturbations in the scalar field are described by,
\begin{equation}
\varphi=\phi(\vec{x},t)-\bar{\phi}(t).
\end{equation}

The joint stress-energy conservation (\ref{eq:conservation:1}) and  (\ref{eq:conservation:2}) leads to the equations of motion for the fluctuations,
\begin{eqnarray}
\dot{\delta}_c+\theta_c+\frac{\dot{h}}{2}&=&\beta\dot{\varphi}\,,
\label{eq:fluctuations:1}\\
\dot{\theta}_c+\mathcal{H}\theta_c &=& \beta\left(k^2\varphi-\dot{\phi}\theta_c\right),\qquad\,
\label{eq:fluctuations:2}\\
\ddot{\varphi}+2\mathcal{H}\dot{\varphi}+\left(k^2+a^2V_{,\phi\phi}\right)\varphi+\frac{\dot{h}\dot{\phi}}{2}&=&-\beta a^2\bar{\rho_c}\delta_c\,,
\label{eq:fluctuations:3}
\end{eqnarray}
in Fourier space, where the dot denotes derivative with respect to conformal time, $\mathcal{H}\equiv aH$ is the conformal Hubble expansion rate, and $\theta_c$ is the velocity divergence of the dark matter fluid with respect to the expansion.

Since the public versions of CLASS do not allow for coupled equations, we modify the perturbation modules of the source code to implement them. We also impose the condition $\theta_c(k,\tau_{ini})=0$ to define the initial timelike hypersurface of the synchronous gauge. However, this condition is insufficient to fix the gauge completely because the coupling makes $\theta_c$ change with time according to Eq.~(\ref{eq:fluctuations:2}). Therefore, we numerically evolve the perturbation equations in a synchronous gauge that still possesses one remaining degree of freedom. The coupling, which can be interpreted as a fifth force, affects the dark matter geodesics. Since the gauge is not comoving with coupled dark matter, it is necessary to further transform $\delta_c$ into the gauge-invariant quantity $\delta_c^C$ that enters the matter density contrast used to predict power spectra that are physically independent from the choice of the gauge. In practice, we modify the existing gauge transformation performed by the public version of the code \cite{Dio_2013} into the following gauge-invariant matter density contrast,
\begin{equation}
\delta_m^C=\frac{\delta\rho_m}{\bar{\rho}_m}+\left(3-\frac{\beta\phi^\prime\bar{\rho_c}}{\bar{\rho_c}+\bar{\rho_b}}\right)\frac{\mathcal{H}}{k^2}\theta_m,
\end{equation}
with $\delta_m\equiv(\delta\rho_b+\delta\rho_c)/(\bar{\rho}_b+\bar{\rho}_c)$. The adiabatic initial condition of dark matter also differs from $\Lambda$CDM. Using the coupled dark matter continuity equation (\ref{eq:cdm}) with the regular photons energy conservation in Eq.~(\ref{eq:radiation}), we obtain that initially,
\begin{equation}
\delta_c=\frac{3}{4}\left(1-\frac{\beta\phi^\prime}{3}\right)\delta_\gamma,
\end{equation}
where $\delta_\gamma$ is the initial photon density contrast. As for quintessence, we adhere to the rationale of the uncoupled case already defined in CLASS that sets vanishing initial perturbations by assuming that the scalar field swiftly reaches the radiation attractor.

\begin{figure}[t]
\includegraphics[scale=0.3]{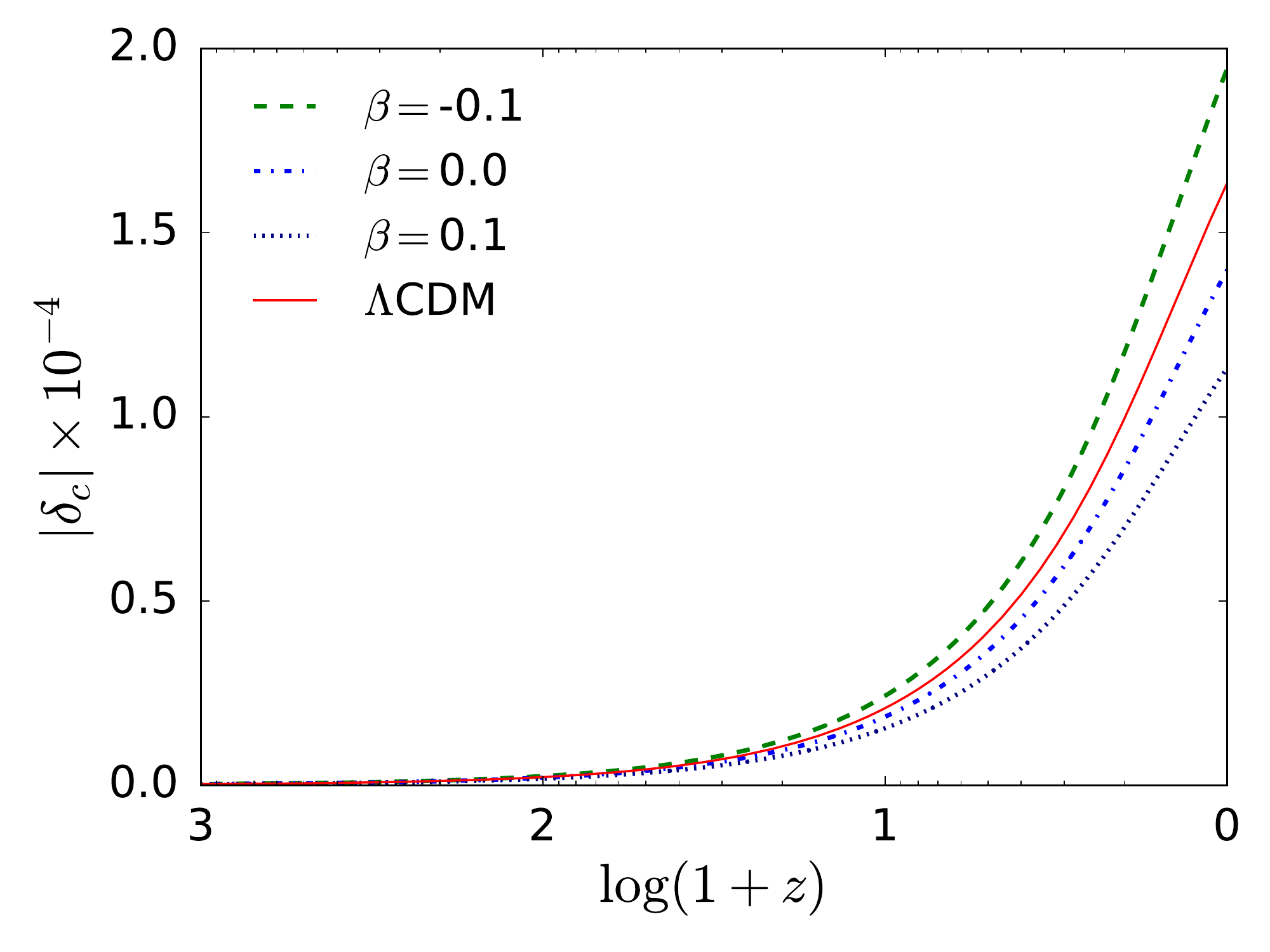}
\includegraphics[scale=0.3]{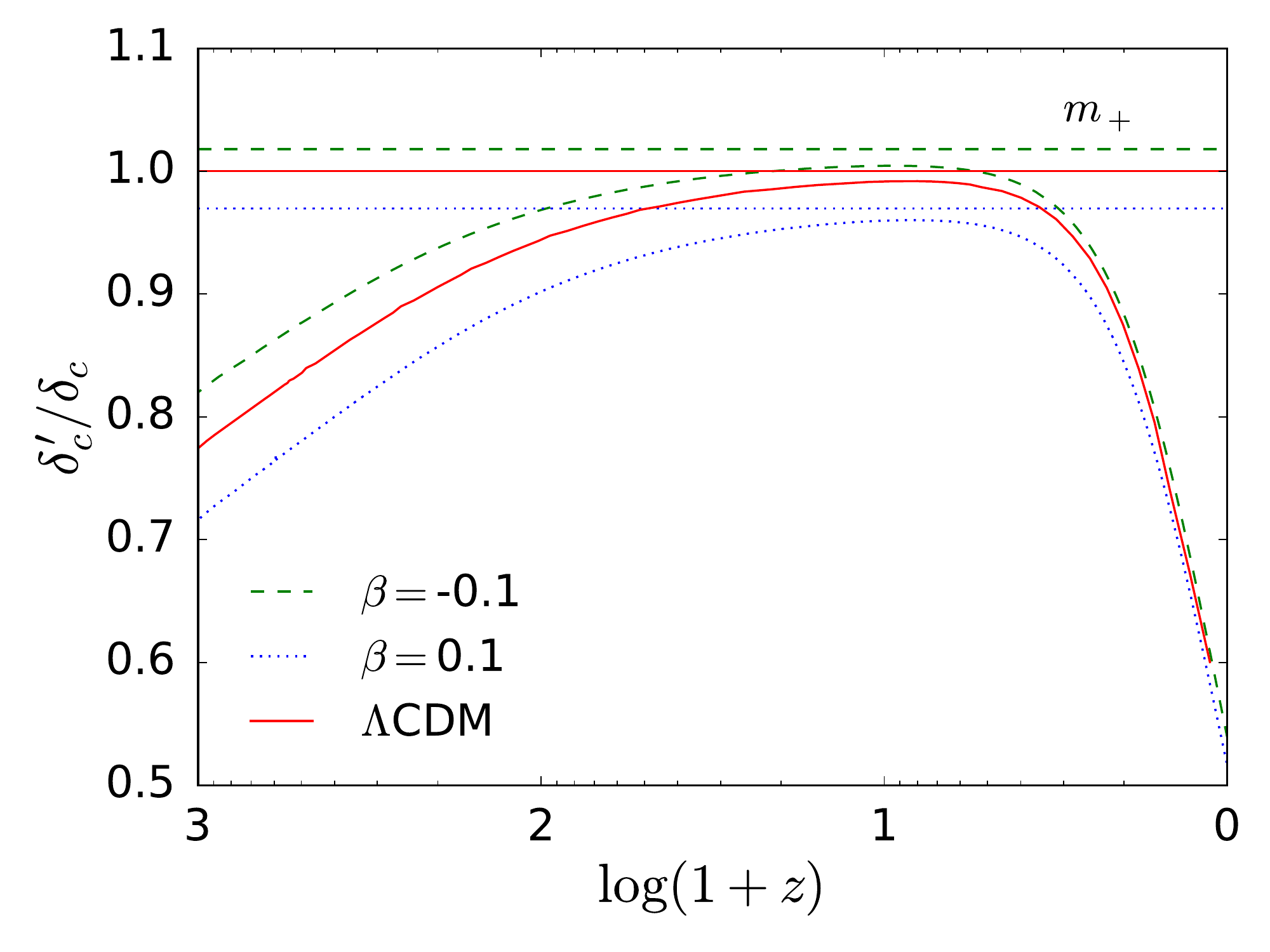}
\caption{\label{fig:f2} The scale considered is $k=0.1~h$/Mpc in the matter era for the instance $\lambda=0.3$. Left panel: evolution of dark matter fluctuations. The initial conditions are adiabatic and the initial spatial curvature perturbation is normalised to $\mathcal{R}=1$. Right panel: evolution of the growth rate function.}
\end{figure}

We validate the code modifications by comparing the numerical and analytic results during the matter dominated epoch when the presence of the coupling makes a difference in the Newtonian limit \cite{Amendola_2002}. For those scales within the horizon we can consider that $k^2\gg\mathcal{H}^2$. Under this approximation, we also neglect radiation and baryons, as well as the higher order terms, i.e. the derivatives of $\varphi$ and the potential term. Being of higher order, we can also neglect the velocity divergence terms in Eq.~(\ref{eq:fluctuations:1}) to approximate $\dot{h}/2=-\dot{\delta}_c$. We combine Eqs.~(\ref{eq:fluctuations:1}), (\ref{eq:fluctuations:2}) and (\ref{eq:fluctuations:3}) with the following equation of motion for the metric trace obtained with the Einstein's field equations,
\begin{equation}
\label{eq:metric_trace}
\ddot{h}+\mathcal{H}\dot{h}+a^2\bar{\rho_c}\delta_c+2\left(2\dot{\phi}\dot{\varphi}-a^2V_{,_\phi}\varphi\right)=0,
\end{equation}
to find the equation of motion for the coupled dark matter perturbations, neglecting higher orders and valid on small scales during the matter dominated era,
\begin{equation}
\ddot{\delta}_c+\left(\mathcal{H}+\beta\dot{\phi}\right)\dot{\delta}_c-\frac{3}{2}\mathcal{H}^2\Omega_c\left(1+2\beta^2\right)\delta_c=0.
\end{equation}
In our parametrisation, where $\phi^\prime=\lambda$, we rewrite the previous equation with the number of e-folds instead of conformal time,
\begin{equation}
\delta^{\prime\prime}_c+\frac{1}{2}\left(1+3\lambda\beta\right)\delta^\prime_c-\frac{3}{2}\left(1-\frac{\lambda^2}{3}-\frac{\lambda\beta}{3}\right)\left(1+2\beta^2\right)\delta_c=0.
\label{eq:newtonian_limit}
\end{equation}
As illustrated by the numerical results in Fig.~\ref{fig:f2}, when energy flows from dark matter to dark energy ($\beta<0$), the growth of matter fluctuations is enhanced against an uncoupled scenario. Three individual effects contribute to this, similarly to Ref.~\cite{PhysRevD.85.023503}. Firstly, the source term is increased by the higher density of dark matter, counterbalancing the effects of early dark energy. Since the interaction accelerates the dilution of dark matter, its fractional energy density is higher in the past for the coupled model in order to evolve to the same abundance today. Secondly, the friction term decreases in line with an effective conformal Hubble rate, $\mathcal{H}_{\rm{eff}}\equiv\mathcal{H}\left(1+\beta\lambda\right)$, that eases the clustering of dark matter particles. Thirdly, the latter is further facilitated by the extra long range gravitational force induced by the coupling that produces an effective potential applying to dark matter particles, $G_{\rm{eff}}\equiv G\left(1+2\beta^2\right)$. On the other flow direction ($\beta>0$) when quintessence is loosing energy, the lower dark matter density at early times combined with the increased expansion rate fight fluctuations growth against the coupling force. It is possible to derive a growing power-law solution to Eq.~(\ref{eq:newtonian_limit}) of the form $\delta_c\propto e^{m_+N}$ in the matter era \cite{Amendola_2002},
\begin{equation}
m_+=\frac{1}{4}\left[-1-3\lambda\beta+\sqrt{24\left(1-\frac{\lambda^2}{3}-\frac{\lambda\beta}{3}\right)\left(1+2\beta^2\right)+\left(1+3\lambda\beta\right)^2}~\right].
\label{eq:growing_m_coupled}
\end{equation}
This analytic expression of the growth rate provides for the validation of the code modifications as it matches well the numerical results under our approximation. It is important to notice that when $\lambda=\beta=0$, the standard $\Lambda$CDM model is retrieved, in which $m_+=1\Rightarrow\delta_c\propto a$.

\begin{figure}[t]
\includegraphics[scale=0.3]{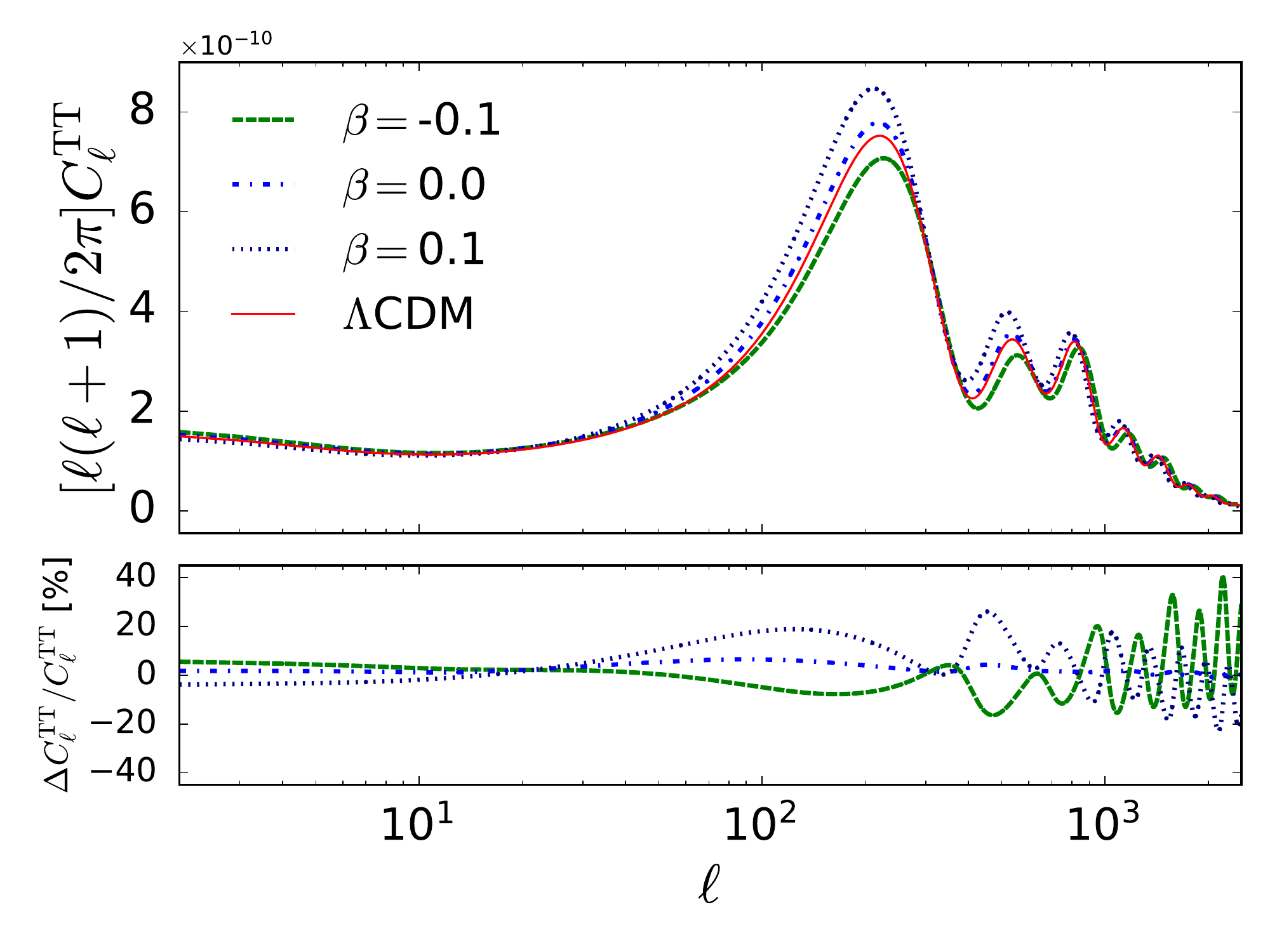}
\includegraphics[scale=0.3]{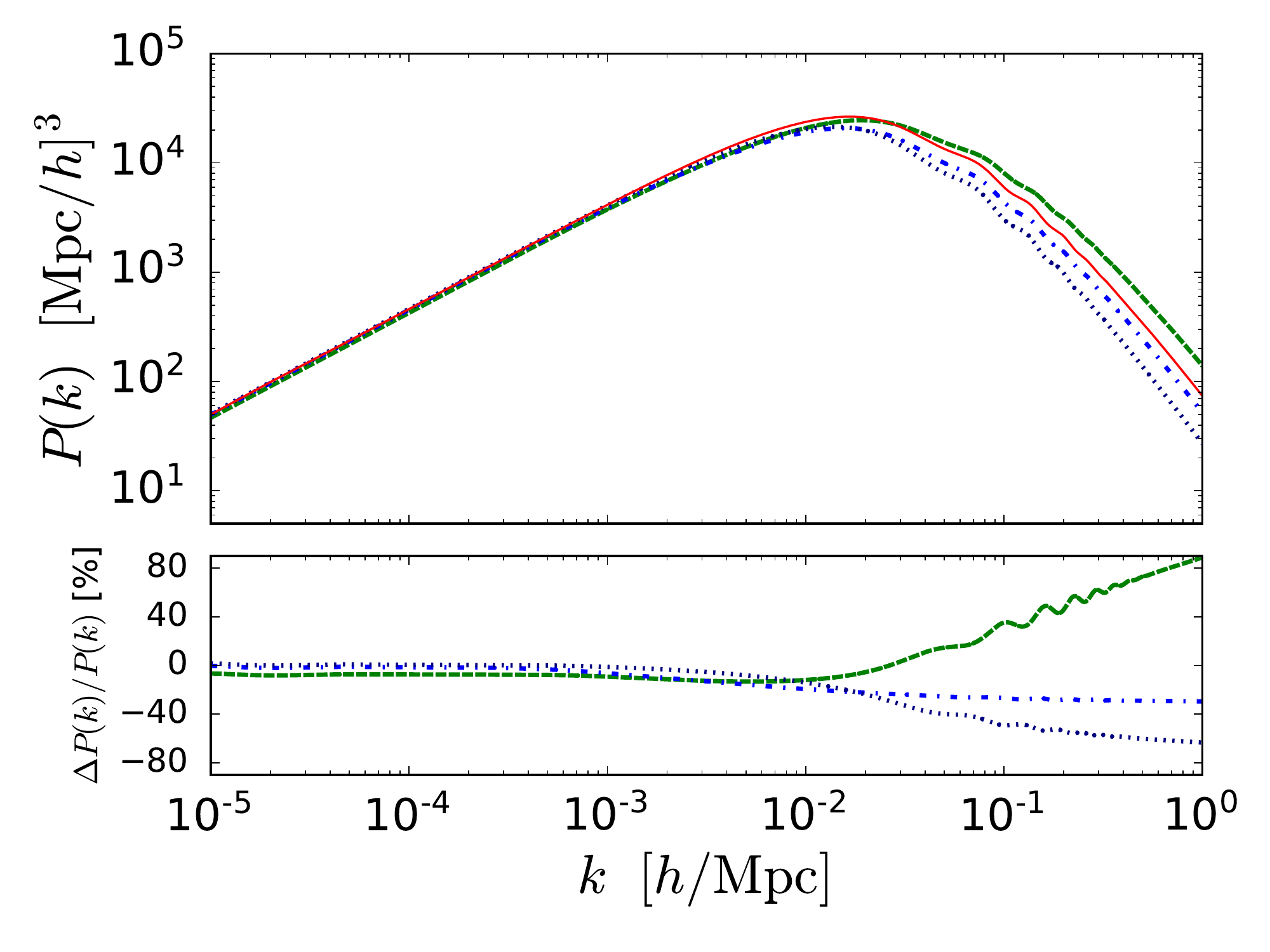}
\caption{\label{fig:power_spectra}  Observables predictions for $\lambda=0.3$. Left panel: CMB dimensionless angular power spectrum. Right panel: linear matter power spectrum at redshift $z=0$.}
\end{figure}

We can now predict the CMB angular power spectrum and the linear matter power spectrum with CLASS, as illustrated in Fig.~\ref{fig:power_spectra}. The combination of diverse effects on the power spectra allows to constrain the parameters with relevant observations \citep{Amendola_2000}. For example, the amplitude of the first peaks in the CMB is influenced by the fractional energy density of matter at the time of photon decoupling which depends on the coupled scalar field. Also, the peak scale is altered compared to $\Lambda$CDM since the interaction modifies the sound horizon at decoupling as well as the comoving angular distance to the last scattering surface. As for the matter power spectrum, the location of the turnover scale is impacted too as it depends on the Hubble radius at matter-radiation equality. For instance, assuming a fiducial cosmology and a negative coupling ($\beta < 0$), the turnover moves to smaller scales when energy is being pumped away from dark matter. As the horizon is shorter at equality, perturbations on smaller scales have time to enter it and grow during radiation domination. Additionally, the coupling accelerates the growth of dark matter fluctuations during the subsequent matter dominated era, further increasing power on small scales. The opposite effects apply to $\beta>0$.
\section{\label{sec:constraints}Observations and model comparison}
We use measurements of the CMB and weak gravitational lensing to fit the scalar field parametrisation, with early and late cosmological probes respectively. The corresponding predicted observables are numerically computed with our modified version of CLASS which is interfaced with the MontePython inference package \cite{MP1,MP2} to extract parameters constraints. Fitting the model consists in estimating the set of parameters that maximises the likelihood, L, i.e. the probability of obtaining the data assuming our model, or equivalently minimises the chi-square, $\chi^2$ .

\begin{figure}[h!]
\centering
\includegraphics[scale=0.8]{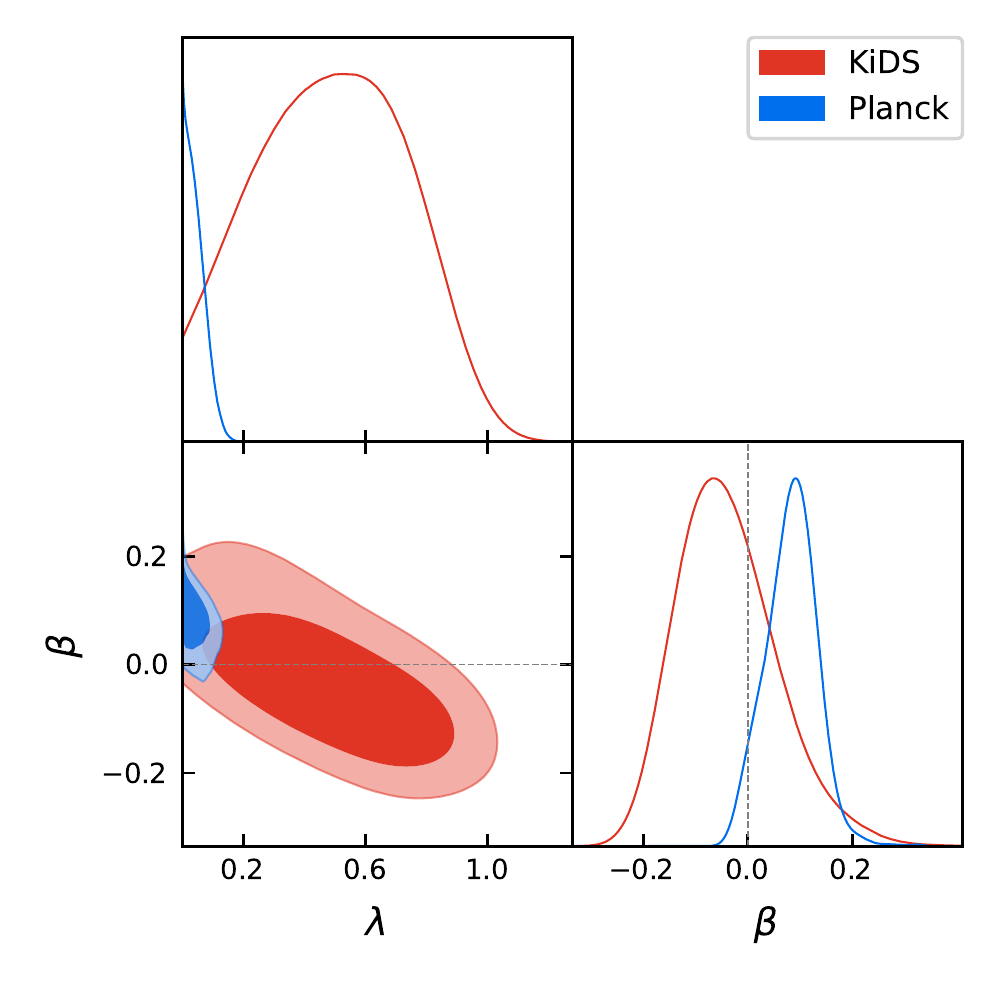}
  \caption{Constraints on the ($\lambda$,$\beta$) plane. Probability distribution and marginalised 2-dimensional correlation (contours include 68$\%$ and 95$\%$ of the probability).}
\label{fig:specific}
\end{figure}

We sample with flat priors the parameter space composed of $\lambda$ and $\beta$, as well as all the usual primary parameters of the flat standard cosmology with adiabatic perturbations power-law spectrum (parametrised by a spectral index $n_s$ and an amplitude $A_s$ normalised at a pivot scale $k_*=0.05$ Mpc$^{-1}$), along with the necessary nuisance parameters that are specific to the experiments.  The Monte Carlo samples are analysed and plotted with the GetDist package \cite{Lewis:2019xzd}. The resulting constraints on the main posteriors are reported in Table~\ref{tab:cosmo}. The posteriors correlation and probability distribution for the specific parameters of the model are depicted in Fig.~\ref{fig:specific}. The results presented in this Section focus on $\lambda>0$. It is worth mentioning that the entire parameter space has been explored, including the symmetric negative values.
\begin{table}[h!]
\centering
\begin{tabular} {l l l}
\hline
Parameter & \multicolumn{1}{c}{Planck}  & \multicolumn{1}{c}{KiDS}\\
\hline
$\lambda $ & $<0.056 $ & $0.50\pm 0.24 $\\ 
$\beta $ & $0.084^{+0.052}_{-0.044} $ & $-0.039^{+0.077}_{-0.11} $\\
$\Omega_m $ & $0.323^{+0.014}_{-0.030} $ & $0.281^{+0.033}_{-0.12} $\\
$H_0 $ & $66.8^{+2.2}_{-1.2} $ & $74.7^{+6.8}_{-2.6} $\\
$\sigma_8 $ & $0.872^{+0.025}_{-0.044} $ & $0.86^{+0.17}_{-0.16} $\\
$S_8 $ & $0.903^{+0.030}_{-0.042} $& $0.784\pm 0.048 $\\
\hline
\end{tabular}
\caption{Constraints obtained on the scalar field parametrisation (mean and 68\% confidence intervals). $S_8\equiv\sigma_8\sqrt{\Omega_m/0.3}$.}
 \label{tab:cosmo}
\end{table}

For the CMB, we choose the 2018 Planck likelihoods \cite{collaboration2019planck}, referred hereafter as "Planck". We restrict ourselves to the temperature auto-correlation $TT$ for lower multipoles (27 data points). Regarding the higher multipoles, we use the lite version of the likelihood, $TT$, $TE$ and $EE$ (613 data points in total) which marginalises over the foreground and instrumental effects, keeping the Planck absolute calibration, $A_{\rm{planck}}$, as the only nuisance parameter. The free parameters in the Bayesian inference are $\{\lambda,\beta,\omega_b,\omega_c,100\theta_s,\ln10^{10}{A_{s}},n_s,\tau_{\rm{reio}},A_{\rm{planck}}\}$, where $\omega_b\equiv\Omega_bh^2$ and $\omega_c\equiv\Omega_ch^2$ are respectively the baryon and cold dark matter physical densities, $h$ being the dimensionless Hubble parameter, $\theta_s$ is the angular size of the sound horizon at the redshift of last scattering, and $\tau_{\rm{reio}}$ is 	the optical depth to reionisation. We use the Metropolis-Hasting sampling algorithm to produce Monte Carlo Markov Chains.

For the weak lensing, we select the Kilo-Degree Survey-450 (KiDS-450) dataset that contains the measurement of galaxies' ellipticity with 450 deg$^2$ imaging (130 data points) \cite{Hildebrandt_2016}. We refer to this dataset as "KiDS". Two nuisance parameters are added in the statistical analysis to account for the measurements' bias: the uncertainty on the amplitude of the Intrinsic Alignment, $A_{\rm{IA}}$, as well as the uncertainty on the dark matter power spectrum amplitude related to the feedback from baryons, $A_{\rm{bary}}$. The free parameters are thus $\{\lambda,\beta,\omega_b,\omega_c,h,\ln10^{10}{A_{s}},n_s,A_{\rm{IA}},A_{\rm{bary}}\}$. We opt for the Nested-Sampling algorithm, implemented as MultiNest in MontePython \cite{Feroz_2008,Feroz_2009,Feroz_2019,Buchner:2014nha}, to sample the parameter space. Moreover, we use in CLASS the HALOFIT model \cite{Takahashi_2012} to reproduce the non-linear scales of structure formation. It is worth noting though that this model might need adaptation to the effective potential applying to dark matter due to the coupling with dark energy.

We further undertake the very same likelihood analysis with the standard $\Lambda$CDM model as a benchmark against which we want to compare the goodness-of-fit of the scalar field parametrisation. The comparison between the two models is summarised in Table \ref{tab:comparison}. The reduced chi-square, $\chi_{\rm{red}}^2\equiv \chi^2/N_{\rm{dof}}$, which is the minimum $\chi^2$ normalised by the number of degrees-of-freedom $N_{\rm{dof}}=N_d-N_{p}$, is a useful criterion to help determine the favoured cosmology. $N_d$ is the number of data points and $N_p$ is the number of free parameters in the Bayesian analysis. The $\chi_{\rm{red}}^2$ test finds that the two models look equivalent for Planck but KiDS slightly prefers the standard one. Diversifying the comparison, similarly to Ref.~\cite{Barros_2020,sagredo2018comparing}, we use another indicator known as the Akaike Information Criterion ($AIC$)~\cite{1100705} that further penalises model complexity.  It can be defined as,
\begin{equation}
AIC=-2\ln L_{\rm{max}}+2N_{p}.
\label{eq:AIC}
\end{equation}
The deviation $\Delta AIC$ from the model minimising the $AIC$, i.e. the best one, is a measure of how the other model is still supported by the observations. It shows that the favourite cosmology for both probes is $\Lambda$CDM which displays the smallest $AIC$. Nonetheless, since the difference is small, the two models can be considered to accurately reproduce Planck observations in light of that indicator. On the other hand, KiDS does show a preference for the standard model, confirming the indication of the reduced chi-square. Since the Nested Sampling algorithm used with KiDS produces the Bayesian evidence for each model \cite{10.1214/06-BA127}, it is possible to compute the Bayesian factor, $B_{\Lambda\phi}$, as the ratio of the evidences, assuming non-committal priors,
\begin{equation}
B_{\Lambda\phi}\equiv\frac{p(d|\Lambda\rm{CDM)}}{p(d|\phi\rm{\,param.})},
\label{eq:bayes_factor}
\end{equation}
where $p(d|\Lambda\rm{CDM)}$ and $p(d|\phi\rm{\,param.})$ are the probabilities of the data given the $\Lambda$CDM model and the scalar field parametrisation respectively. The Bayes factor is a good criterion for models comparison as it uses the evidence which strikes a balance between best-fit and  complexity. Adopting the so-called Jeffreys' empirical scale \cite{Trotta_2008} that categorises the level of preference for a model against another, we find with KiDS a weak evidence for the concordance model over the scalar field parametrisation since $\ln{B_{\Lambda\phi}}=2.02$. 
\begin{table}[tb]
\centering
\begin{tabular} {l c c c c}
\hline
 & \multicolumn{2}{c}{Planck}  & \multicolumn{2}{c}{KiDS}\\
  & \multicolumn{1}{c}{$\Lambda$CDM}  & \multicolumn{1}{c}{$\phi$ param.}   & \multicolumn{1}{c}{$\Lambda$CDM}  & \multicolumn{1}{c}{$\phi$ param.}\\
\hline
$\chi^2$ & $602.6$ & $599.7$ & $321.5$ & $321.7$ \\ 
$\chi^2_{\rm{red}}$ & $1.0$ & $1.0$ & $2.6$ & $2.7$ \\
\hline
$AIC$ & $616.6$ & $617.7$ & $335.5$ & $339.7$ \\
$\Delta AIC$ & $0$ & $1.1$ &  $0$ & $4.2$ \\
\hline
\end{tabular}
\caption{Comparison between the standard $\Lambda$CDM model and the scalar field parametrisation.}
 \label{tab:comparison}
\end{table}
\section{\label{sec:discussion}Analysis of the results and discussion}
According to the Bayesian inference carried out in Section \ref{sec:constraints}, the two sets of observations bring different constraints on our two parameters. The Planck based parameters extraction provides an upper bound for the posterior $\lambda$ which is compatible with a cosmological constant, $\lambda < 0.056$, while the coupling posterior, $\beta = 0.084^{+0.052}_{-0.044}$, corresponds to energy injection from dark energy to dark matter. Conversely, the analysis with the KiDS measurements indicates a preference for rather high values of the scalar field parameter, with $\lambda=0.50\pm 0.24$, yet compatible with the more relaxed ceilings imposed on early dark energy at BBN discussed in Section~\ref{sec:parametrisation}. As regards the coupling posterior, KiDS slightly favours energy injection from dark matter to dark energy with $\beta=-0.039^{+0.077}_{-0.11}$. As already mentioned, the HALOFIT model is not optimised for the existence of the coupling. By removing the 100 data points of the angular scales in the KiDS sample that are sensitive to non-linear physics \cite{10.1093/mnras/stx998}, we have also checked that an analysis restricted to the linear regime of the matter power spectrum does not significantly affect the results.

\begin{figure}[h!]
\centering
\includegraphics[scale=0.55]{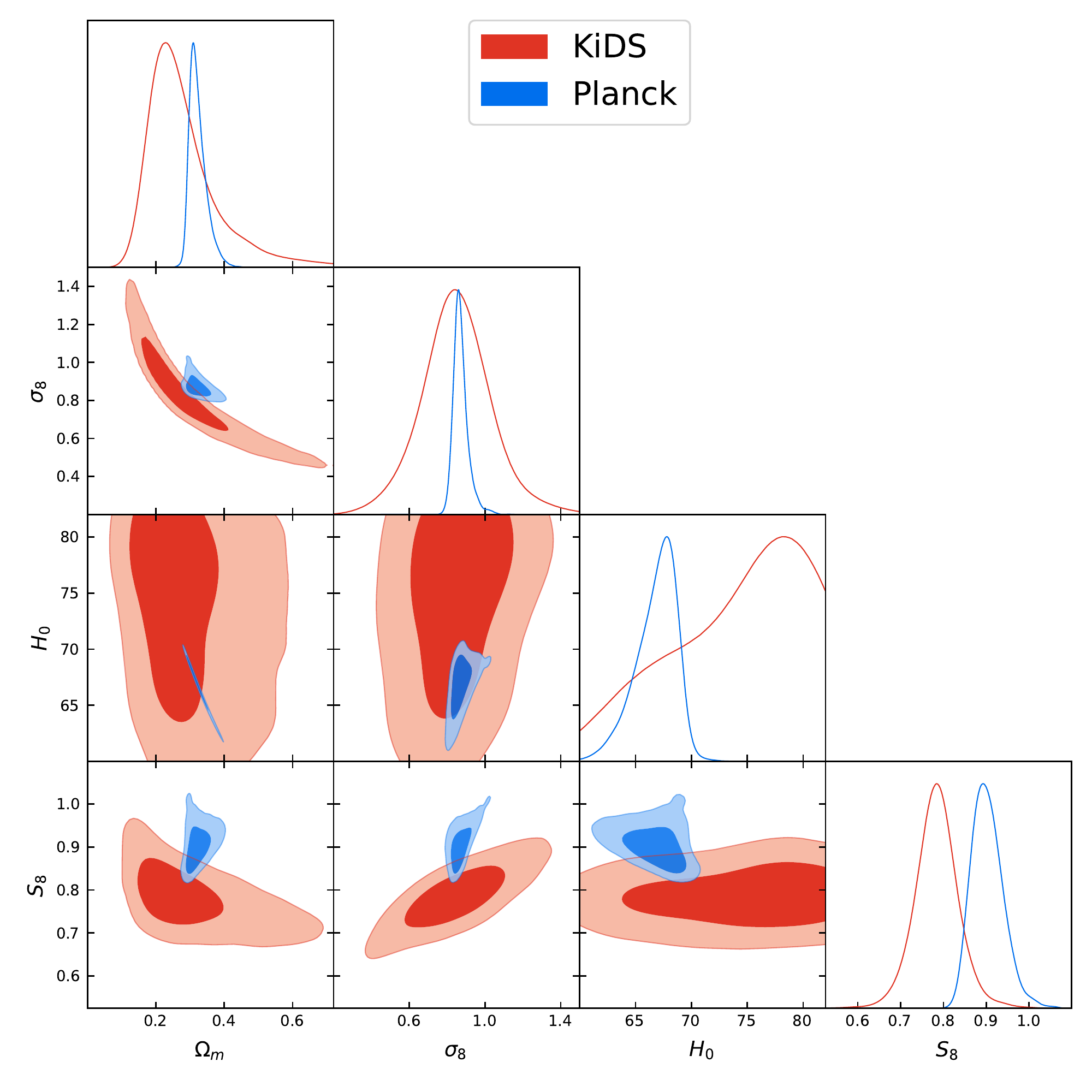}
  \caption{Constraints extracted for the parametrisation of the scalar field in the parameter space. Probability distribution and marginalised 2-dimensional correlation. Contours include 68$\%$ and 95$\%$ of the probability.}
\label{fig:derived}
\end{figure}

The overlapping probability contours in Fig.~\ref{fig:derived} suggest a possible region of the parameter space that allows the model to fit the observations of both Planck and KiDS, despite existing tensions on $\lambda$ (1.9$\sigma$) and $\beta$ (1.1$\sigma$) between the two cosmological probes. Looking at the rate of clustering, the presence of the coupling has the effect of enhancing dark matter fluctuations, leading to increased values of the matter power spectrum amplitude at the scale $8h^{-1}$Mpc, $\sigma_8$, that are preferred by both experiments in our analysis. The parametrisation also seems to decrease the existing standard model tension on $\sigma_8$ between the early and the late Universe, yet at the expense of the creation of tensions on $\lambda$ and $\beta$, preventing us to draw any further conclusions in this regard.

To summarise, we find that despite its simplicity, our parametrisation reproduces the evolution of the background and large scale structure formation. It is also capable of generating the anisotropies of the cosmic microwave background. The difference with the standard model is that dark energy is not negligible in primordial times, although in quantity bound by the Big Bang nucleosynthesis. The merit of this parametrisation consists in the dynamics of the scalar field that unavoidably leads the Universe towards its current acceleration for a wide range of initial conditions. It particularly captures many possible dynamics for dark energy at late times, while keeping its equation of state bound at high redshift. In this respect, our parametrisation is appropriate to help constrain the evolution of the dark energy equation of state as one single additional parameter suffice at low redshift. We therefore conclude that it constitutes a credible and competitive cosmological model although the approach is purely phenomenological. We have also performed this analysis in combination with SnIa and BAO datasets. However the inclusion of the background data does not provide stronger constraints and therefore has not been included in this article. Forthcoming Euclid data, however, will certainly improve the constrains on these kind of models \cite{refId0}, provided that the modelling at non-linear scales can be appropriately optimised.
\section*{Acknowledgements}
The authors thank Giuseppe Fanizza and Miguel Zumalacarregui for their useful comments. This research was supported by Fundação para a Ciência e a Tecnologia (FCT) through the research grants: UIDB/04434/2020 \& UIDP/04434/2020, CERN/FIS-PAR/0037/2019, PTDC/FIS-OUT/29048/2017, COMPETE2020: POCI-01-0145-FEDER-028987 (cosmoESPRESSO), FCT: PTDC/FIS-AST/28987/2017 (DarkRipple).

\bibliography{vdf}

\end{document}